\documentclass[11pt]{article}
\usepackage{moriond,epsfig,amsmath}

\newcommand{\ga}{\gamma}

\newcommand{\om}{\omega}
\newcommand{\as}{\alpha_{\mathrm{s}}}

\def\nc{N_C}
\def\nf{n_f}

\newcommand{\ie}{\textit{i.e.}\ }
\newcommand{\eg}{\textit{e.g.}\ }

\newcommand\hepph[1]{hep-ph/#1}

\newcommand\jhep[3]{{{\it JHEP }{\bf #1} (#2) #3}}

\newcommand\npb[3]{{{\it Nucl. Phys. }{\bf B #1} (#2) #3}}

\newcommand\plb[3]{{{\it Phys. Lett. }{\bf B #1} (#2) #3}}

\newcommand\prd[3]{{{\it Phys. Rev. }{\bf D #1} (#2) #3}}

\newcommand\sjnp[3]{{{\it Sov. J. Nucl. Phys. }{\bf #1} (#2) #3}}
\newcommand\jetp[3]{{{\it Sov. Phys. JETP }{\bf #1} (#2) #3}}

\newcommand\jetpl[3]{{{\it JETP Lett. }{\bf #1} (#2) #3}}

\begin{document}
\vspace*{4cm} 

\title{The status of NLL BFKL
\footnote{Talk presented at
    the XXXVth Rencontres de Moriond, QCD and High Energy Hadronic
    Interactions, Les Arcs, France, March 18-25, 2000.}
}

\author{ G.P. Salam }

\address{TH Division, CERN, CH-1211 Gen\`eve 23.}

\begin{flushright}
  \vspace{-5cm}
  CERN--TH/2000--150\\
  hep-ph/0005304\\
  May 2000\vspace{3cm}
\end{flushright}

\maketitle\abstracts{This talk summarises the current status of the
  NLL corrections to BFKL physics and discusses the question of
  small-$x$ factorisation.}

\section{Introduction and total cross sections}

Understanding perturbative high-energy hadronic scattering is one of the
fundamental problems of QCD, and one which has seen considerable
progress in the past couple of years.

The reaction under consideration is that of the scattering of objects
with some transverse scale $Q^2$ at a centre-of-mass energy
$\sqrt{s}$, in the limit $s \gg Q^2 \gg \Lambda^2_\mathrm{QCD}$. This
was first examined in the mid 1970's with the resummation of the
leading logarithmic (LL) terms $(\as \ln s)^n$ to give the result that
the cross section $\sigma \sim s^{4\ln2 \as \nc/\pi}$. This is known
as the BFKL pomeron.\cite{BFKL} This result however seems to be in
contradiction with numerous experimental results, which invariably
indicate that while there is a rise of the cross section at large $s$,
the power is somewhat lower, of the order of $0.3$ as opposed to the
$0.5$ expected from the LL calculation (for $\as \simeq 0.2$).

It was expected that this discrepancy would be resolved by the
inclusion of the next-to-leading (NLL) corrections $\as (\as \ln
s)^n$, but it turns out that they modify the power $\om$ as
follows:\,\cite{NLL}
\begin{equation}
  \om = 2.65 \as -   16.3\as^2 \simeq -0.12 \qquad \mbox{(for $\as =
    0.2$)}\,,
\end{equation}
which is just as incompatible with the data as the LL result. Closer
inspection reveals that the NLL corrections, taken literally, also
imply negative cross sections for the scattering of two objects of
different transverse sizes\,\cite{Ross} (essentially, in such a collinear 
limit the convergence becomes even worse).

There have been several significantly different attempts to explain
the large size of the corrections and to thus estimate yet
higher-order corrections with the aim of obtaining stable predictions.
One involves the idea of a minimum rapidity gap between
emissions,\cite{Schmidt} while another argues that a meaningful answer
can be obtained using a more `natural' renormalisation scheme coupled
with BLM resummation.\cite{BFKLP} Both suffer from instabilities (with
respect to the choice of the rapidity gap and the renormalisation
scheme, respectively) and have difficulty solving the problem of the
negative cross sections, essentially because they do not take into
account the worsening of the convergence in the collinear limit. 

\begin{figure}[tbp]
  \begin{center}
    \epsfig{file=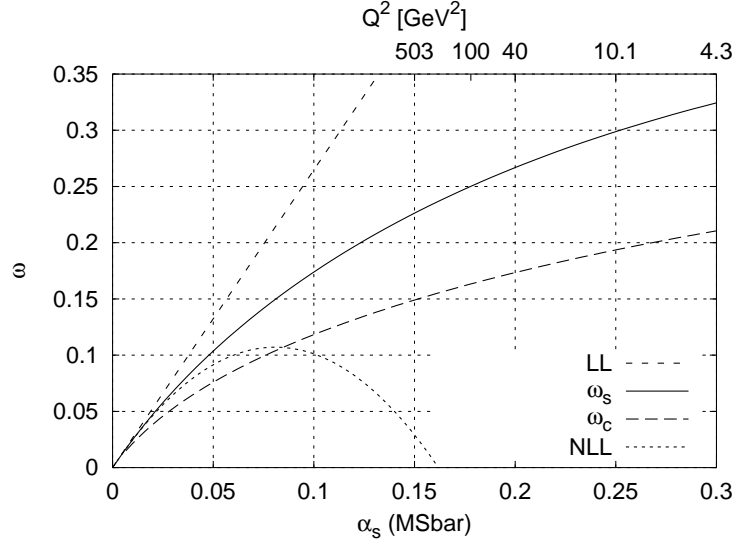, width=0.6\textwidth}
    \caption{The prediction for the BFKL power as a function of $\as$
      at LL, NLL and after the all-order inclusion of collinearly
      enhanced terms ($\om_s$). The uncertainty on $\om_s$ is of the
      order of $10\%$. The curve labelled $\om_c$ is the BFKL power
      for splitting functions. The results, shown for $\nf=0$, depend
      only weakly on $\nf$.}
    \label{fig:omcrit}
  \end{center}
\end{figure}

An alternative approach indeed starts from a consideration of the
collinear limit, and observes that a large part of the NLL corrections
(even for non-collinear scattering) come precisely from terms that are
enhanced in the collinear limit. These can be resummed by requiring
that the cross section satisfy the renormalisation group properties in
both collinear limits (there are two collinear limits corresponding to
which of the scattering objects is the smaller).\cite{CollResum} One
finds that the problem of negative cross sections goes away completely
and that the power of the cross section growth is relatively stable
with respect changes of scheme. The result for the power as a function
of $\as$ is shown in fig.~\ref{fig:omcrit}.

The cleanest (most inclusive) measurement of the BFKL power comes from
$\ga^*\ga^*$ scattering, by the L3 collaboration\,\cite{L3} at scales
$Q^2=3.5$ and $14.5$~GeV$^2$.  They quote a value for the power of
$\om = 0.37\pm 0.04$. The quoted error is probably an underestimate:
in the fit procedure the normalisation is fixed to be the LL value,
whereas one should fit also for the normalisation because it too is
liable to be subject to significant (but as yet uncalculated)
higher-order corrections. A more realistic error is about three times
larger. There are also systematic errors associated with the
functional form used to fit for the power.  So the measurement is
roughly in accord with the theoretical expectation which goes from
$0.29$ to $0.33$ for this range of scales (for comparison the LL power
is in the range $0.6$ to $0.8$), but a satisfactory comparison
requires, on the experimental front, higher precision data and larger
energies (\eg at the NLC), and on the theoretical front a calculation
of the expected normalisation (\ie the impact factors to NLL order).

\section{Scaling violations}

So far we have discussed the total cross section in a context where
both scattering objects are perturbative. An important situation is
that in which only one of the objects is perturbative, namely deep
inelastic scattering. Since the other object is non-perturbative the
calculation of the total cross section is beyond perturbation theory.
When the $s$ is of the same order as $Q^2$ (\ie $x$ not too small) we
know however that the cross section can be factorised into
non-perturbative parton distributions convoluted with perturbative
coefficient functions, and the scaling violations of the parton
distributions can be calculated perturbatively through the DGLAP
equations.\cite{DGLAP} The intuitive justification for this picture is
that emissions are ordered in transverse scale, so that the
non-perturbative parton distribution contains emissions only below a
given scale $Q_0^2$, while the coefficient function (and parton
distribution evolution) involves only subsequent emissions, which are
all above $Q_0^2$.

However in high-energy scattering (\ie at small $x$) it has long been
known that emissions are not ordered in transverse momentum. This led
to the suggestion\,\cite{Mueller} that the factorisation which holds at
moderate $x$ might break down at small $x$. (Recently, on this basis,
it has been argued that small-$x$ splitting functions are beyond
calculation and can at best be fitted.\cite{ABF})

Last summer an explicit counterexample was presented,\cite{CCS2}
namely a toy model which retained the fundamental property of BFKL
evolution, namely emissions which are not ordered in transverse scale,
but in which the property of factorisation could be demonstrated for
all $x$. Recently, numerical methods have been developed\,\cite{CCST}
which allow the study of factorisation in the full BFKL equation (for
the time being at LL, with running coupling). In this approach one
solves the BFKL equation to obtain a gluon distribution as a function
of $x$ and $Q^2$. This distribution depends intrinsically on the
regularisation of $\as$ in the infra-red (IR) and the initial
transverse scale. One then performs a deconvolution of the scaling
violations from the gluon distribution, to obtain an effective
splitting function.
\begin{figure}[tbp]
  \begin{center}
    \epsfig{file=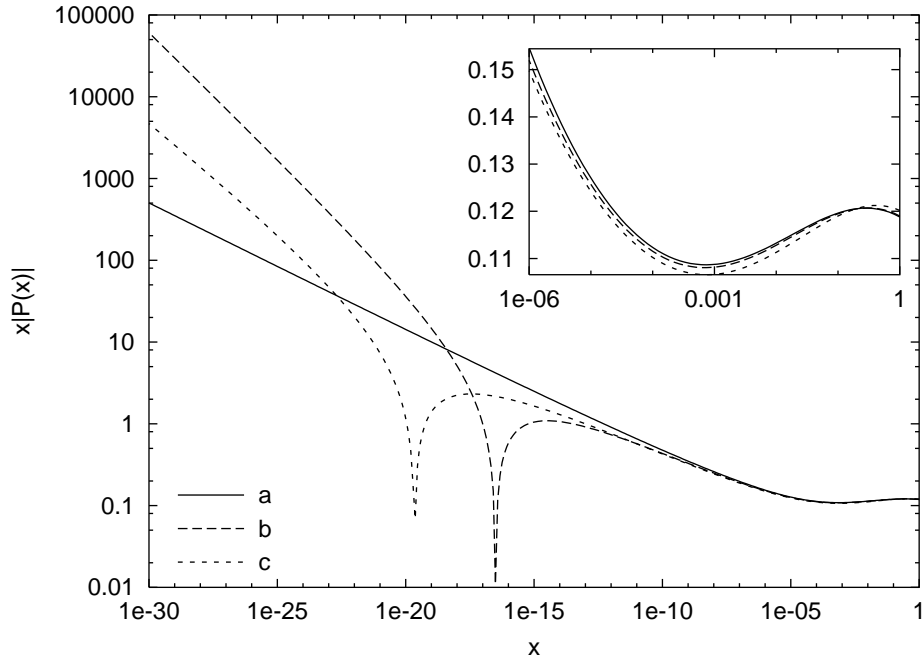,width=0.8\textwidth}
    \caption{The splitting function in LL BFKL with running coupling
      for three different sets (a,b,c) of infra-red regularisation and
      initial transverse scale. (Shown for $\as=0.126$.) The inset
      shows the same curves in a more limited range of $x$ values.}
    \label{fig:split}
  \end{center}
\end{figure}

The results of this procedure are shown in fig.~\ref{fig:split}, which
shows the effective splitting function for three different sets of
infra-red regularisations. One sees that the splitting function is
independent of the IR regularisation, except at very small $x$, where
a non-perturbative component takes over. More detailed investigation
reveals that this second component is higher-twist, and despite its
much stronger $x$ dependence, the scaling violations themselves can be
predicted to within a relative higher-twist term which is not
$x$ enhanced, thus confirming that factorisation holds even at small
$x$. 

A rough explanation of why factorisation holds is the following: the
fact that the evolution is not ordered in transverse momentum means
that the separation into a parton distribution and coefficient or
splitting functions is less trivial. If one chooses to separate the
two at some scale $Q_0$, then the parton distribution now depends on
emissions both below and above $Q_0^2$. But significantly, the
splitting and coefficient functions turn out to still only depend on
emissions above $Q_0^2$.

This brings us to the question of what exactly the small-$x$ splitting
functions look like. A fundamental property is that the power-growth
only sets in at relatively small $x$. It is also significantly weaker
than what would be observed in total cross sections at the same scale.
This is the reason for the presence of two powers in
fig.~\ref{fig:omcrit}: $\om_s$ is that which applies to total cross
sections, while $\om_c$ is that which is relevant for splitting
functions. These two points (which have been noted also by
Thorne\,\cite{Thorne}) mean that for the $x$ values that are relevant at
HERA, the small-$x$ resummed splitting functions are probably quite
similar to the DGLAP splitting functions, explaining the unexpected
success of the latter in reproducing the observed scaling violations.

\section*{Acknowledgements}
I wish to thank Marcello Ciafaloni and Dimitri Colferai (in collaboration
with whom many of the results presented here have been obtained) for
discussions. This work was supported in part by the E.U. QCDNET contract
FMRX-CT98-0194.


\section*{References}
\begin{thebibliography}{99}
\bibitem{BFKL} L.N. Lipatov, \sjnp{23}{1976}{338};
       E.A. Kuraev, L.N. Lipatov and V.S. Fadin, \jetp{44}{1976}{443};
       E.A. Kuraev, L.N. Lipatov and V.S. Fadin, \jetp{45}{1977}{199};
       Ya. Balitskii and L.N. Lipatov, \sjnp{28}{1978}{822}.

\bibitem{NLL} V.S. Fadin and L.N. Lipatov, \plb{429}{1998}{127}
  [\hepph{9802290}];   M. Ciafaloni and G. Camici,
  \plb{430}{1998}{349} [\hepph{9803389}]; references therein.

\bibitem{Ross} D.A. Ross, \plb{431}{1998}{161} [\hepph{9804332}].

\bibitem{Schmidt} C. Schmidt, \prd{60}{1999}{074003} [\hepph{9901397}].
  
\bibitem{BFKLP} S.J. Brodsky, V.S. Fadin, V.T. Kim, L.N. Lipatov and
  G.B. Pivovarov, \jetpl{70}{1999}{155} [\hepph{9901229}].

\bibitem{CollResum} G.P. Salam, \jhep{07}{1998}{19} [\hepph{9806482}];
  M. Ciafaloni and D. Colferai, \plb{452}{1999}{372}
  [\hepph{9812366}]; M. Ciafaloni, D. Colferai and G.P. Salam,
  \prd{60}{1999}{114036} [\hepph{9905566}].  

\bibitem{L3} L3
  Collaboration (M. Acciarri et al.), \plb{453}{1999}{333}; C.H. LIN,
  these proceedings.

\bibitem{DGLAP} V.N. Gribov and L.N. Lipatov, \sjnp{15}{1972}{438};\\
  G. Altarelli and G. Parisi, \npb{126}{1977}{298};\\
  Yu.L. Dokshitzer, \jetp{46}{1977}{641}.

\bibitem{Mueller} A.H. Mueller, \plb{396}{1997}{251} [hep-ph/9612251].
  
\bibitem{ABF} G. Altarelli, R.D. Ball and S. Forte, hep-ph/9911273.

\bibitem{CCS2} M. Ciafaloni, D. Colferai and G.P. Salam,
  \jhep{10}{1999}{017} [\hepph{9907409}].  

\bibitem{CCST} M. Ciafaloni, D. Colferai, G.P. Salam and M. Taiuti, in
  preparation.

\bibitem{Thorne} R.S. Thorne, \plb{474}{2000}{372} [\hepph{9912284}].
\end{thebibliography}
\end{document}